\documentclass[aps,prstab,twocolumn,amsmath,amssymb,groupedaddress]{revtex4-1}

\usepackage{siunitx}
\usepackage{placeins}
\usepackage{float}
\usepackage[colorlinks=true]{hyperref}
\hypersetup{ 
    citecolor=black,%
    filecolor=black,%
    linkcolor=black,%
    urlcolor=black 
}
\usepackage{graphicx}
\usepackage{dcolumn}
\usepackage{bm}
\usepackage{color}

\begin{document}

\title{Longitudinal phase space synthesis with tailored 3D-printable dielectric-lined waveguides}

\author{F.~Mayet}
\email{frank.mayet@desy.de}
\affiliation{DESY, Notkestrasse 85, 22607 Hamburg, Germany}
\author{R.~Assmann}
\affiliation{DESY, Notkestrasse 85, 22607 Hamburg, Germany}
\author{F.~Lemery}
\email{francois.lemery@desy.de}
\affiliation{DESY, Notkestrasse 85, 22607 Hamburg, Germany}

\date{\today}

\begin{abstract}
Longitudinal phase space manipulation is a critical and necessary component for advanced acceleration concepts, radiation sources and improving performances of X-ray free electron lasers.  
Here we present a simple and versatile method to semi-arbitrarily shape the longitudinal phase space of a charged bunch by using wakefields generated in tailored dielectric-lined waveguides. We apply the concept in simulation and provide examples for radiation generation and bunch compression. We finally discuss the manufacturing capabilities of a modern 3D printer and investigate how printing limitations, as well as the shape of the input LPS affect the performance of the device.
\end{abstract}

\maketitle

\section{Introduction}
Emerging advanced accelerator concepts require precise control over the longitudinal phase space (LPS) of charged particle beams. Efficient beam-driven acceleration, for example, relies on longitudinally-tailored electron bunch profiles which can be produced with an appropriate energy modulation and dispersive section \cite{englandPRL, piotPRL, lemeryPRAB, andonianPRAB}. Phase-space linearization for bunch compression is especially important to optimize the performance of multistage linacs and X-ray free electron lasers (XFELs) \cite{paoloLinear, emmaLinear, antipovLinear}. There are several ways to control the LPS. Energy modulation approaches via self-wakes in e.g. dielectric or corrugated structures provide attractive and simple methods to produce microbunch trains and large peak currents \cite{antipovTHz, lemeryPRAB, lemeryPRL}. Laser-based energy modulation techniques using magnetic chicanes are particularly useful for FEL seeding \cite{hghg, eehg1, eehg2, eehgDemo1, eehgDemo2} and for beam acceleration \cite{sudar1}.  Arbitrary laser-based phase space control was discussed in \cite{Hemsing:2013vv}, illustrating the potential for producing different current profiles for various applications. Unfortunately however, although the scheme works well in simulation, the approach is complex to implement, requiring several undulators and magnetic chicanes in addition to the modulating laser.

In this paper we explore arbitrary waveform synthesis using self-wakes produced in dielectric-lined waveguides (DLW). By using segmented waveguides with varying cross sections, the excited wakefields carry different spectral contents throughout the structure, enabling control over the energy modulation across the bunch. The dependence of the modal content on the DLW geometry allows for enough degrees of freedom to optimize such a segmented structure according to the desired output LPS. Due to the nature of the physical process, the scheme is completely passive, removing the need for synchronization with e.g. a modulating laser beam. In the following, the device is referred to as a longitudinal phase space shaper (LPSS).

The paper is structured as follows: Section~\ref{sec:DLW_Wakefields} provides an overview on 1D wakefield theory, Section~\ref{sec:Fourier} discusses Fourier synthesis for single-mode structures, Section~\ref{sec:Multimode_Example} provides examples for multimode structure optimizations using computational optimization, Section~\ref{sec:RealisticStructures} discusses the impact of manufacturing limitations of modern 3D printers by investigating the effect of printing resolution and segment transitions on the excited wakefields. Finally, Section~\ref{sec:robustness} discusses the effect of slight variations in the shape of the input LPS on a figure of merit of an output LPS, based on an example optimization case.
\section{Wakefield generation in a DLW} \label{sec:DLW_Wakefields}
\begin{figure}[hbp]
  \centering
  \includegraphics[width=0.45\columnwidth]{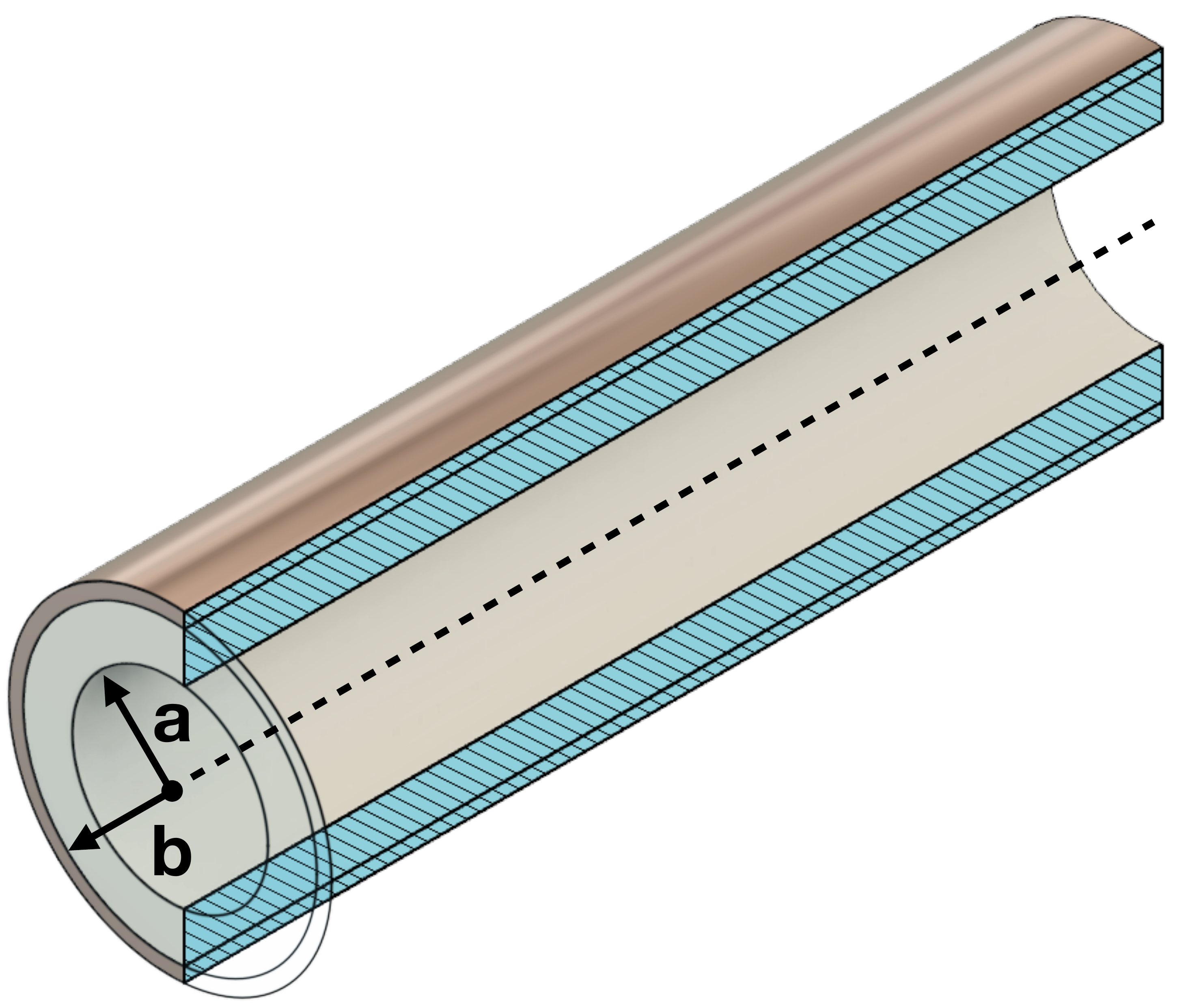}
  \caption{Schematic of a cylindrical dielectric-lined waveguide. The lining with dielectric constant $\epsilon_r$ has an inner radius $a$ and an outer radius $b$. It is coated with a thin metallic layer on the outside.}
  \label{fig:03_DLW_Schematic}
\end{figure}

The theory of Cherenkov wakefield generation in cylindrically-symmetric DLWs is well described in \cite{vossWeiland, ROSING:1990gj, ng}. Here we follow \cite{ROSING:1990gj}, for a structure with inner radius $r=a$, outer radius $r=b$ and dielectric permittivity $\epsilon_r$. The outer surface is assumed to be coated with a perfect conductor. See Fig.~\ref{fig:03_DLW_Schematic} for a schematic. A more rigorous theoretical investigation could include conductive and dielectric losses in DLWs \cite{dlwMikhail, klausDLW}.

\begin{figure}[htbp]
  \centering
  \includegraphics[width=\columnwidth]{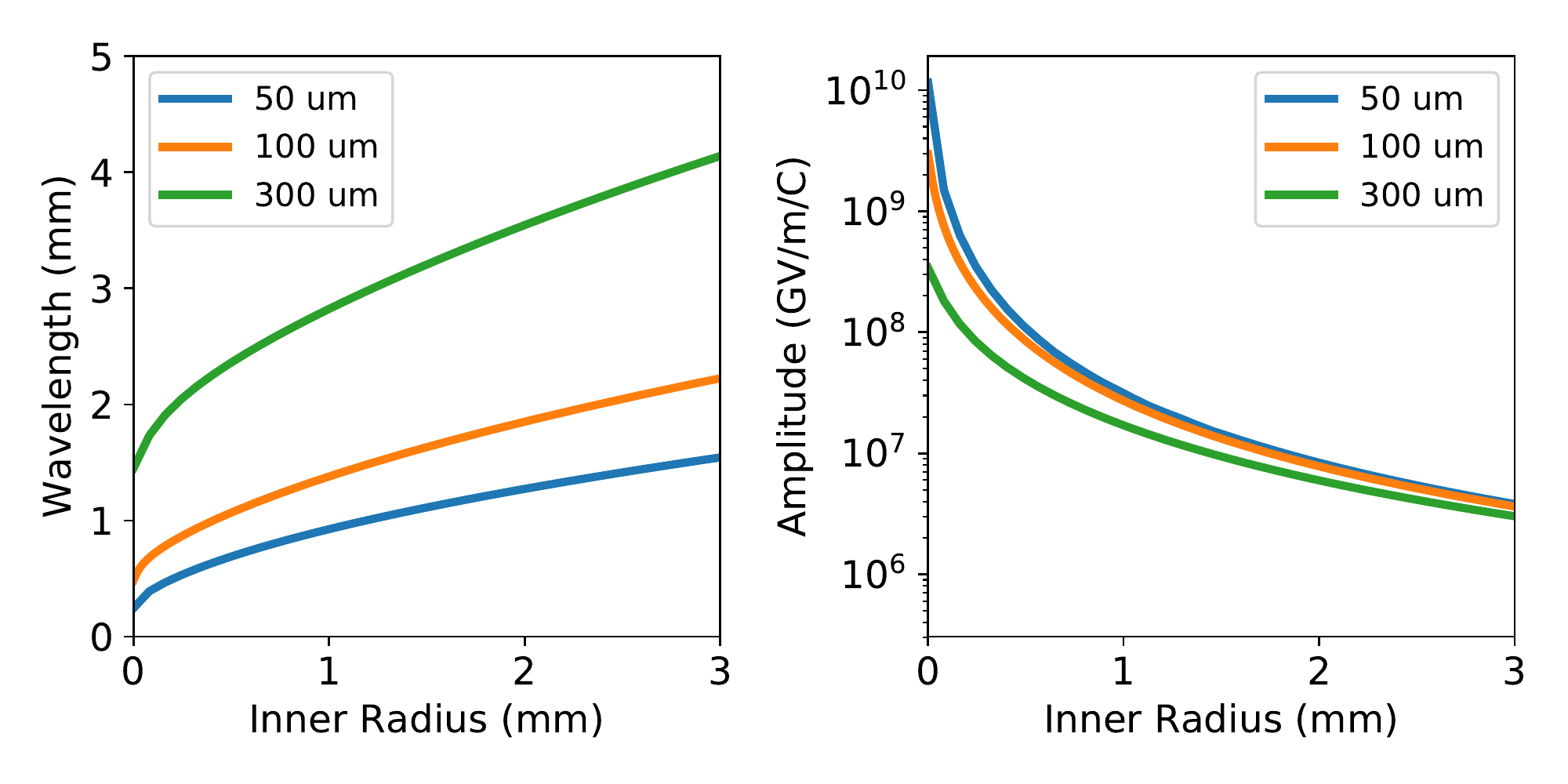}
  \caption{Plot of the numerically calculated wavelength and amplitude of a wakefield excited by an on-axis electron bunch in a single-mode DLW. The different colors correspond to different thicknesses of the dielectric lining.}
  \label{fig:DLW_Modes_Geometry}
\end{figure}

In the ultrarelativistic limit, a point charge travelling on-axis ($r=0$) will excite a wakefield with a corresponding Green's function with $M$ modes \cite{Chao:1993pc, Stupakov:2000sp},
\begin{equation}
    G(t)=\sum_{m=1}^M \kappa _m \cdot \cos(2 \pi f_m t),
\end{equation}
where $\kappa_m$ and $k_m$ are the loss factor and wave number of the $m$th mode respectively and are calculated numerically \cite{ROSING:1990gj, piotGitWake}. This Green's function is often also referred to as the single particle wake potential $W_z(\tau)\,[V/C]$, where $\tau$ denotes the time difference between the point charge and a trailing witness charge. Note that it is defined by the boundary conditions and hence - in our case - the geometry of the DLW. By varying e.g. the inner radius $a$ of a DLW, it can be seen that both wavelength and amplitude depend strongly on the geometry of the structure (see Fig~\ref{fig:DLW_Modes_Geometry}). Considering that the amplitude of the longitudinal wakefield scales as $1/a^2$ \cite{ROSING:1990gj}, it becomes apparent that potentially very high field strengths can be reached in small aperture DLWs.

The overall wake potential $V(t)$ produced by a bunch can be calculated by convolving its current profile $I(t)$ with $W_z(\tau)$. Therefore
\begin{equation}
	 V(t) = -\int_{-\infty}^t I(\tau) W_z(t-\tau)d\tau.
	 \label{eq:03_half_convo}
\end{equation}
The field excitation can also be described in terms of the frequency dependent bunch form factor $F$. Then
\begin{equation}
	V(t)=q \cdot \sum_{m=1}^M F_m\kappa _m \cdot \cos(2 \pi f_m t),
\end{equation}
where $q$ is the total charge of the bunch. A strong mode excitation therefore requires a bunch with an appropriate spectral content i.e. a relatively short bunch, or also by having a relatively short rise time in e.g. a flat-top distribution \cite{pitzLaser,lemeryPRL}.

In a cascaded arrangement of multiple DLWs, outside of experimental constraints due to e.g. limitations in beam transport, the energy modulations via wakefields from different structures can be concatenated. The following section illustrates the broad potential for a set of cascaded, or a single segmented structure to produce a versatile range of energy modulations. We note that the usage of segmented structures, and the produced effects of transient wakes is discussed in Section~\ref{sec:RealisticStructures}.
\section{LPS Shaping in Single-Mode Structures}\label{sec:Fourier}
Fourier synthesis provides a simple way to produce a large variety of waveforms which have various applications in conventional electronics. Here we explore how Fourier synthesis can be applied to charged particle beams using self-wakes imparted in high-impedance mediums, e.g. DLWs. We are specifically interested in the Fourier series for odd functions, since the wakefield at the head of the bunch must be zero.

In the simplest case, each of the individual segments of an LPSS is a single mode structure with a specific fundamental mode frequency and amplitude. As discussed above, the wake function $W_z(\tau)$ for such a structure is simply given by
\begin{equation}
    W_{z,m}(\tau)=\kappa _m \cdot \cos(2\pi f_m \tau).
\end{equation}
Using this and Eq.~\ref{eq:03_half_convo}, the energy modulation imparted by a single DLW segment $n$ can be estimated as (cf. \cite{Lemery:2014prstab})
\begin{equation}
    \begin{split}
    \Delta E_{n}(t) = &- l_n\cdot \kappa _{m(n)}\\&\cdot \int_{-\infty}^t I(\tau) \cos(2\pi f_{m(n)} (t-\tau))d\tau,
    \end{split}
\end{equation}
where $l_n$ is the length of the $n$th DLW segment. The total energy modulation imparted by an $N$-segment structure can hence be estimated as
\begin{equation}
    \Delta E_\text{tot}(t) = \sum_{n=0}^N  \Delta E_{n}(t)
\end{equation}
(see Section~\ref{sec:RealisticStructures} for a discussion on the effects of sharp segment transitions on the resulting wakefields). Assuming an idealized flat-top current profile $I(\tau)$, the total energy modulation reduces to
\begin{equation}
    \Delta E_\text{tot}(t) = \sum_{n=0}^N A _n\cdot \sin(2\pi f_{m(n)} t),
    \label{eq:idealdeltaEtot}
\end{equation}
where $A_n$ is the amplitude factor of the $n$th segment. Considering the scaling laws shown in Fig.~\ref{fig:DLW_Modes_Geometry}, arbitrary LPS shapes can be constructed via Fourier composition. The amplitude $A_n$ of each frequency component can be adjusted by choosing an appropriate $l_n$. It should be noted that the harmonic content of the input current profile must be sufficient to excite the desired modes. 

\begin{figure}[htbp]
  \centering
  \includegraphics[width=0.9\columnwidth]{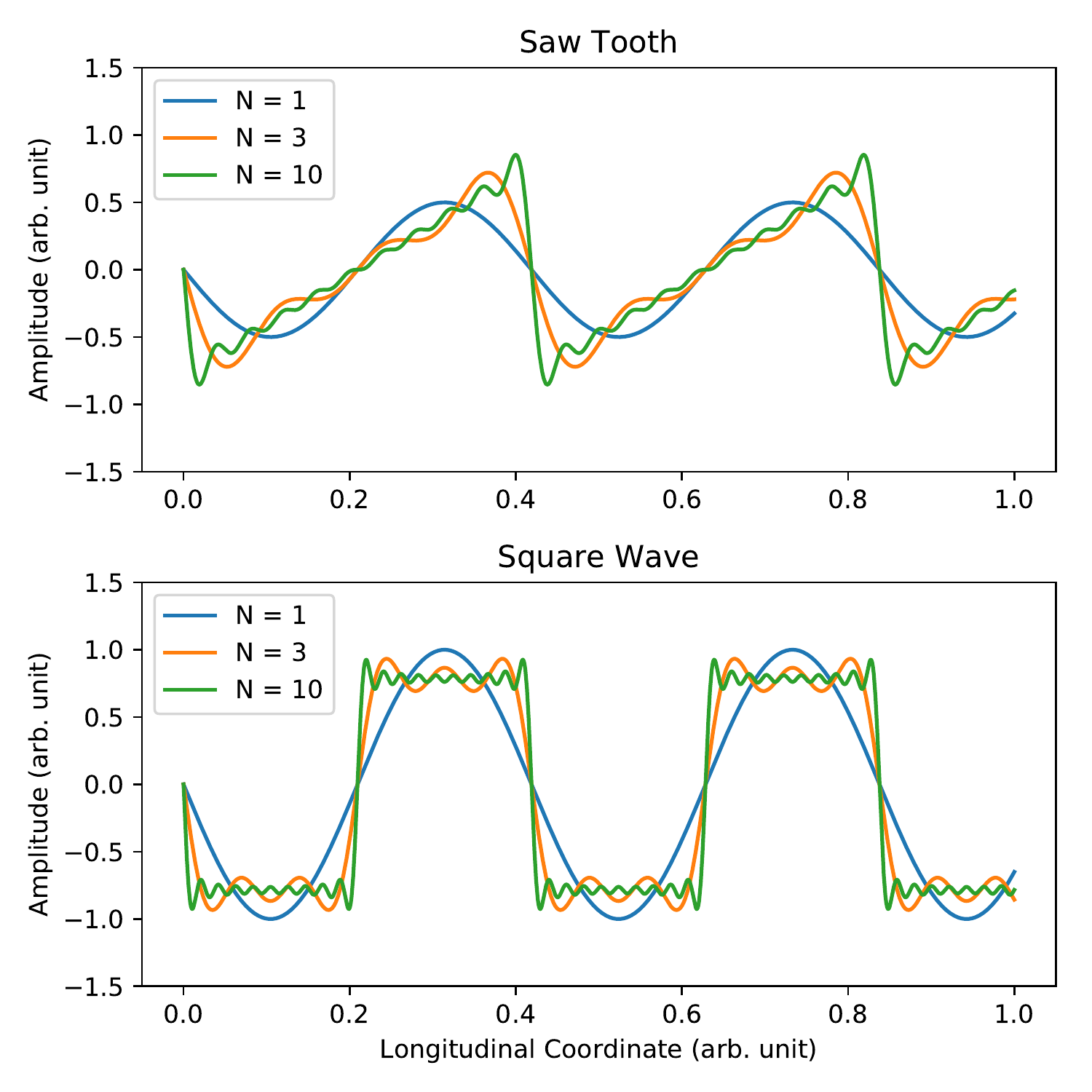}
  \caption{Plot of amplitude vs. long. coordinate for an arbitrary saw-tooth modulation (Eq.~\ref{eq:sawtoothdefinition}) and an arbitrary square wave modulation (Eq.~\ref{eq:squarewavedefinition}) for $N=1$, $N=3$ and $N=10$.}
  \label{fig:modulationtypes}
\end{figure}

Eq.~\ref{eq:idealdeltaEtot} essentially corresponds to an ordinary Fourier sine series. A saw-tooth wave, for example, can be constructed by summing up only even harmonics with proper normalization. Hence, the Fourier series for a given fundamental frequency $f_0$ is given by
\begin{equation}
    F_\text{saw}(t) = A\cdot \sum_{n=0}^\infty \frac{1}{2n + 2} \sin(\pi (2n+2)f_0 \cdot t),
    \label{eq:sawtoothdefinition}
\end{equation}
where $A$ is an amplitude scaling factor. Another simple example is a square wave. Its Fourier series only contains odd harmonics. Thus
\begin{equation}
    F_\text{squ}(t) = A\cdot \sum_{n=0}^\infty \frac{1}{2n + 1} \sin(2\pi (2n+1)f_0 \cdot t).
    \label{eq:squarewavedefinition}
\end{equation}
Fig.~\ref{fig:modulationtypes} visualizes the two modulation types for different values of $N$.

\begin{figure*}[htbp]
	\centering
	\includegraphics[width=0.94\textwidth]{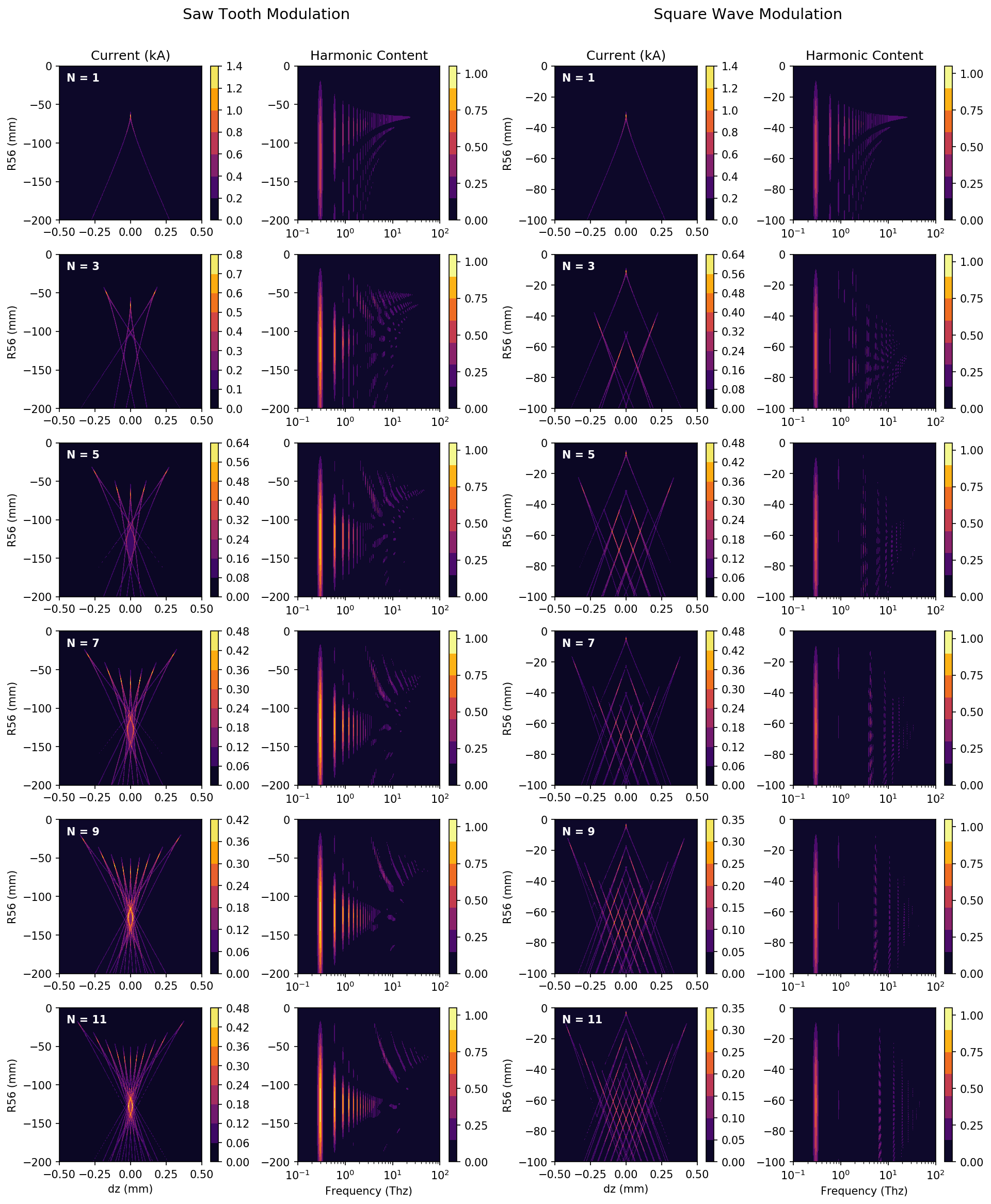}
	\caption{Contour plots of both the beam current within a single fundamental modulation wavelength $\lambda _0 = \SI{1}{\milli\meter}$, as well as the harmonic content of the bunch vs. the longitudinal dispersion $R_\text{56}$. The scan was performed for $N \in [1,3,5,7,9,11]$. Both a saw-tooth modulation according to Eq.~\ref{eq:sawtoothdefinition}, as well as a square wave modulation according to Eq.~\ref{eq:squarewavedefinition} are shown. The idealized input current is assumed to be of flat-top shape and the initial energy $E_0 = \SI{100}{\mega e\volt}$ is constant along the bunch. It has a total bunch length of \SI{1}{\milli\meter} and $Q = \SI{42}{\pico\coulomb}$. The assumed maximum modulation depth of the lowest frequency component is \SI{500}{\kilo e\volt}. Note that a high slice energy spread would lead to blurring out the small features in the respective phase spaces. Here we assume a cold beam in order to explore the full mathematical potential of the scheme.}
	\label{fig:singleModeGallery}
\end{figure*}

In order to explore possible use cases of such energy modulations we investigated the effect of applying linear longitudinal dispersion ($R_\text{56}$) to the the phase space. In this work we adopt the convention that the head of the bunch is at $z<0$ and $R_{56} < 0$. Fig.~\ref{fig:singleModeGallery} shows contour plots of both the beam current within a single fundamental modulation wavelength $\lambda _0 = \SI{1}{\milli\meter}$, as well as the harmonic content of the bunch vs. the longitudinal dispersion $R_\text{56}$ for different values of $N$. The idealized input current is assumed to be flat-top. We also assume a cold beam in order to be able to explore the full mathematical limits of the scheme. The investigation is carried out for both a saw-tooth modulation (cf. Eq.~\ref{eq:sawtoothdefinition}), as well as for a square wave modulation (cf. Eq.~\ref{eq:squarewavedefinition}). It can be seen, as longitudinal dispersion is applied, interesting features emerge. 

In the case of the saw-tooth modulation first the higher frequency modulation on the rising part of the saw-tooth (see Fig.~\ref{fig:modulationtypes}) is compressed. Then, as $R_\text{56}$ increases, the minimum and maximum of the saw-tooth converge, which results in a current spike. The current spike is more defined as $N$ increases, which can be attributed to a less pronounced Gibbs ringing at the sharp edges of the saw-tooth, as well as an overall flattening for higher values of $N$. This behaviour is also represented by the ellipsoidal shape visible in the contour plots of the beam current vs. $\text{d}z$ and $R_{56}$, which becomes narrower as $N$ increases (see Fig~\ref{fig:singleModeGallery}). It is interesting to note that as the amplitude of the high frequency modulation along the rising part of the saw-tooth varies, different parts of the rising edge require different values of $R_\text{56}$ for optimal compression. This is clearly visible in the contour plots. For symmetry reasons, always two sub-microbunches emerge. By adjusting $R_\text{56}$, a specific pair of microbunches with a defined relative distance can be selected. It has to be noted, however, that - depending on the modulation depth - these sub-structures require very low slice energy spread to be significant vs. the background. If the respective bunching factor $b_n$ should not be reduced by more than roughly a factor of 2, then $\delta_\text{mod}/\delta_\text{sl} \leq n$ has to be satisfied, where $n$ is the harmonic number of $f_0$ and $\delta_\text{mod}$ and $\delta_\text{sl}$ are the relative modulation depth and slice energy spread respectively; cf. \cite{Hemsing:2014ju}. 

In case of the square wave modulation the plots show a different behaviour. As $R_{56}$ increases, first a single current spike is formed, which corresponds to the sharp edge of the energy modulation. As the edge becomes sharper (higher $N$), optimal bunching occurs for smaller values of $R_{56}$. Increasing $R_{56}$ beyond optimal bunching reveals a very particular rhombus pattern in the contour plot, which is explained by the fact that the negative and positive plateaus of the square wave are shifted on top of each other. The higher the value of $N$, the more intricate the rhombus pattern becomes. It is interesting to note that - by applying appropriately high $R_{56}$ - the two plateaus of the square wave modulation will form two sub-microbunch trains at their own distinct energy levels ($E_0 \pm \Delta E$).

\begin{figure}[htbp]
	\centering
	\includegraphics[width=\columnwidth]{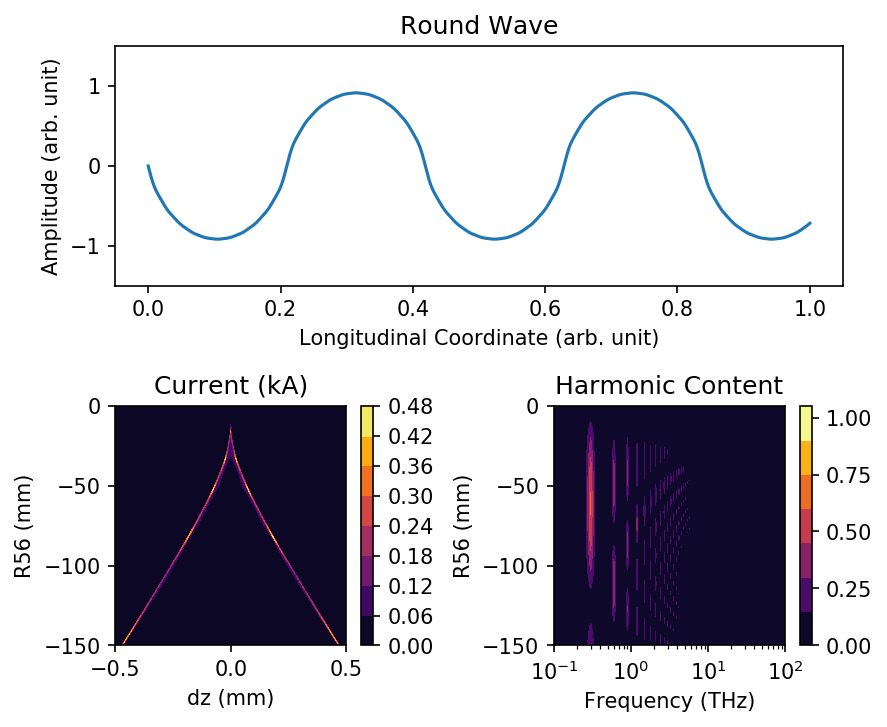}
	\caption{\textbf{Top:} Plot of an $N=11$ round wave modulation according to Eq.~\ref{eq:roundwavedefinition}. \textbf{Bottom:} Contour plots of both the beam current within a single fundamental modulation wavelength $\lambda _0 = \SI{1}{\milli\meter}$, as well as the harmonic content of the bunch vs. the longitudinal dispersion $R_\text{56}$. The scan was performed for $N = 11$. The idealized input current and modulation depth is assumed to be the same as described in Fig.~\ref{fig:singleModeGallery}.}
	\label{fig:singleModeRndWaveExample}
\end{figure}

The saw-tooth and square wave modulation are only two examples of possible Fourier series based LPS modulations. Many other interesting waveforms might exist, which are not discussed here. In order to show how drastic even small changes to a particular Fourier series definition can be, one can consider squaring the normalization factor in Eq.~\ref{eq:squarewavedefinition}. This yields, instead of a sharp square wave, a smooth rounded wave. The definition now reads
\begin{equation}
    F_\text{rnd}(t) = A\cdot \sum_{n=0}^\infty \frac{1}{(2n + 1)^2} \sin(2\pi (2n+1)f_0 \cdot t).
    \label{eq:roundwavedefinition}
\end{equation}
Fig.~\ref{fig:singleModeRndWaveExample} shows both the shape of an $N = 11$ round wave modulation, as well as contour plots analogous to Fig.~\ref{fig:singleModeGallery}. It can be seen that applying $R_{56}$ to this kind of modulation at first glance leads to a dependence similar to a simple sine wave modulation. The main difference, however, is that the beam current of the sub-microbunches, which occur after over-bunching, shows multiple additional maxima of similar magnitude compared to the initial single microbunch. In the case of a simple sine modulation the peak current would decrease rapidly. As the number of additional maxima increases with $N$, this means that using a high-$N$ round wave modulation, one can obtain high-quality sub-microbunches with semi-continuously adjustable relative spacing.
\section{Arbitrary Multimode Optimization} \label{sec:Multimode_Example}
So far we have investigated Fourier shaping of an idealized input LPS. In order to work with arbitrary input distributions, a more sophisticated optimization routine must be used. This is especially true if multi-mode DLW segments are to be included, as the number of degrees of freedom gets too large for manual optimization. Hence a routine based for example on the particle swarm algorithm (PSO) must be employed \cite{Kennedy:2004wp}. The algorithm varies all geometric parameters of the individual segments at the same time in order to find a global minimum of a given merit function. This merit function is given in the LPSS case by the similarity of the resulting LPS to the desired LPS shape. Since segment radius, length and wall thickness can be varied, the resulting number of independent variables is $3N$, where $N$ is the number of segments. For the LPSS study presented here, the PSO was implemented using PyOpt \footnote{\url{http://www.pyopt.org}, last access: 23rd of April 2019.}. At each iteration step either a simulation using a specifically generated input file for a numerical simulation code, or a semi-analytical calculation based on Eq.~\ref{eq:03_half_convo} is carried out. If space-charge effects are neglected, the difference between the numerical simulation using ASTRA \cite{ASTRAASpaceChar:LTSRiAsm} and the semi-analytical approach was found to be negligible. Hence, the much faster semi-analytical calculation was used for the simulations shown in the following discussion. 

\begin{figure*}[htbp]
	\centering
	\includegraphics[width=\textwidth]{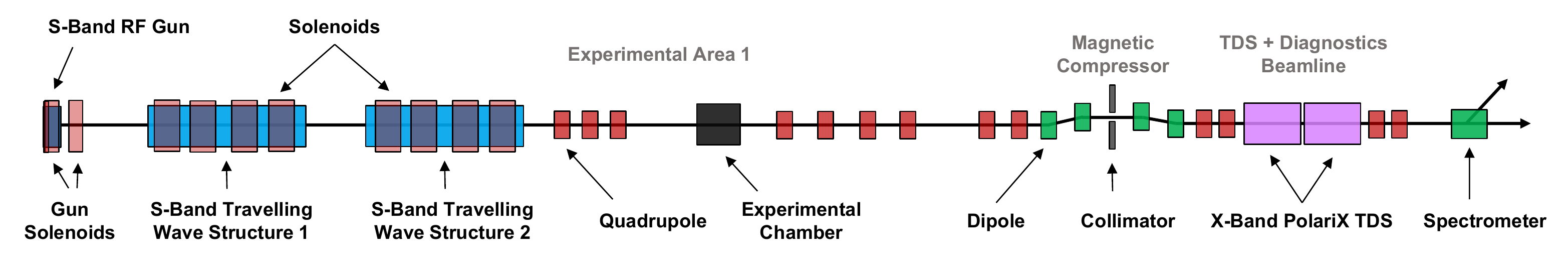}
	\caption{Schematic of the layout of the ARES linac at DESY (not to scale). The LPSS interaction is simulated to take place in the experimental chamber of \emph{Experimental Area 1} at $z=\SI{16.8}{\meter}$.}
	\label{fig:ARESLinac}
\end{figure*}
As an example optimization goal the linearization of an incoming LPS obtained from close to on-crest acceleration was chosen. This scenario is interesting, because the resulting LPS shows a clear signature of the sinusoidal RF field of the linac structures, which would limit the achievable bunch length in subsequent compression. In order to keep the number of free parameters manageable, the number of LPSS segments was limited to 10. The optimizer was configured to bring the Pearson's $R$ value of centered $\tilde n\sigma_z$ regions within the final distribution as close to 1 as possible. Here $\tilde n \in \mathbb{N}$. Table~\ref{tab:LPSS_Opt_Settings} summarizes the possible ranges of values for the geometry parameters of the 10 individual segments.
\begin{table}[!htb]
   \centering
   \caption{LPSS optimization variable ranges for each of the 10 segments.}
   \begin{ruledtabular}
   \begin{tabular}{lc}
   \textbf{Parameter} & \textbf{Value} \\
   \hline
   Inner radius & $[0.1, 2.5]\,\SI{}{\milli\meter}$\\
   Dielectric thickness & $[50, 1000]\,\SI{}{\micro\meter}$\\
   Segment length & $[1, 100]\,\SI{}{\milli\meter}$\\
   \end{tabular}
   \end{ruledtabular}
   \label{tab:LPSS_Opt_Settings}
\end{table}

For the input we consider three different electron bunch distributions with \SI{10}{\pico\coulomb}, \SI{100}{\pico\coulomb} and \SI{200}{\pico\coulomb} of total charge and a mean energy of $\sim \SI{100}{\mega e\volt}$, based on numerical simulations of the ARES linac at DESY \cite{Marchetti:2020EAAC}. This is done in order to provide a realistic example, which could be used as the basis for future experimental verification of the scheme. Fig.~\ref{fig:ARESLinac} shows a schematic of the ARES lattice. If both linac structures are driven at their respective maximum gradients of $\sim\SI{25}{\mega\volt /\meter}$, a final mean energy up to $\sim \SI{150}{\mega e\volt}$ is possible. The decision to limit the example working points to $\sim \SI{100}{\mega e\volt}$ is a practical one, as the overall LPSS structure length generally increases with the required LPS modulation strength and the experimental chamber at ARES imposes strict space limitations. The three working points were optimized to minimize transverse emittance at the interaction point ($z=\SI{16.8}{\meter}$) for three different charges using ASTRA, including space charge effects. Table~\ref{tab:ARES_WPs} summarizes the respective beam parameters.
\begin{table}[!htbp]
   \centering
   \caption{Beam parameters of the three ARES linac working points (WP) at the interaction point ($z=\SI{16.8}{\meter}$), obtained from ASTRA simulations. Initial spatial and temporal profile: Gaussian. TWS: Travelling Wave Structure.}
   \begin{ruledtabular}
   \begin{tabular}{lccc}
   \textbf{Parameter} & \textbf{WP1} & \textbf{WP2} & \textbf{WP3} \\
   \hline
   Charge & \SI{10}{\pico\coulomb} & \SI{100}{\pico\coulomb} & \SI{200}{\pico\coulomb}\\
   TWS injection phase & \SI{-3}{\degree} & \SI{-5}{\degree} & \SI{-8}{\degree}\\
   $E_0$ & \SI{108}{\mega e\volt} &  \SI{109}{\mega e\volt} &  \SI{108}{\mega e\volt}\\
   $\sigma _E / E_0$ & $2.8 \cdot 10^{-4}$ &  $4.1 \cdot 10^{-3}$ &  $5.3 \cdot 10^{-3}$\\
   $\sigma _t$ & \SI{673}{\femto\second} & \SI{1.95}{\pico\second} & \SI{2.65}{\pico\second}\\
   $\varepsilon _{\text{n},x,y}$ & \SI{146}{\nano\meter} & \SI{370}{\nano\meter} & \SI{465}{\nano\meter}\\
   \end{tabular}
   \end{ruledtabular}
   \label{tab:ARES_WPs}
\end{table}

\begin{figure}[H]
	\centering
	\includegraphics[width=\columnwidth]{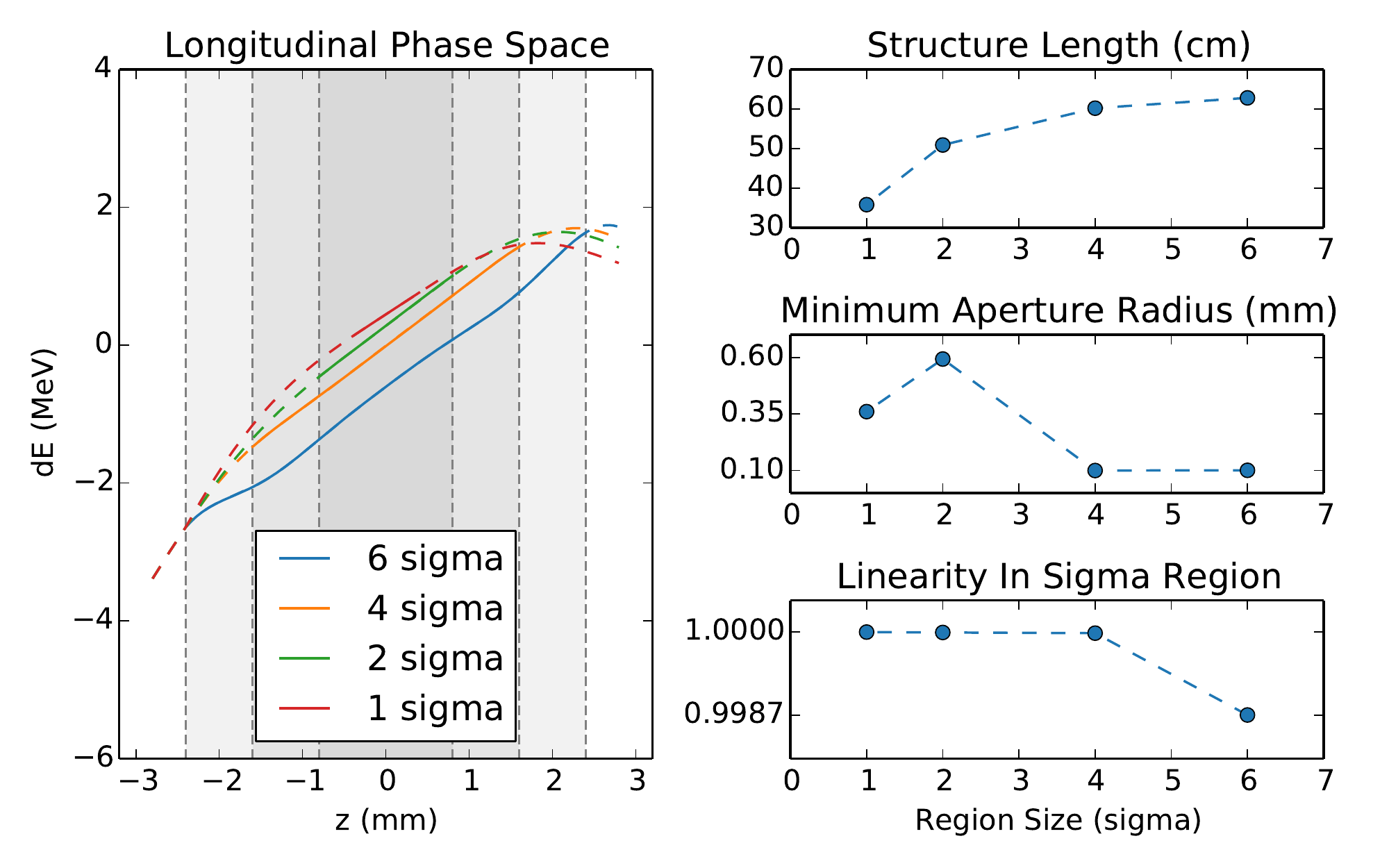}
	\caption{\textbf{Left}: Comparison of LPSS linearization results, depending on the size of a defined region of interest (ROI) within the bunch. The solid part of the lines corresponds to the respective ROI. \textbf{Right:} Total LPSS structure length, minimal segment aperture radius within the structure and linearity of the output LPS within the respective ROI, depending on the ROI size.}
	\label{fig:LPSS_Sigma_ROI_Comp}
\end{figure}

\newpage
In order to first investigate the effect of limiting the optimization goal to specific $\tilde n\sigma_z$ regions within the LPS on the resulting LPSS geometry, four different optimization runs were performed. As input, WP3 was chosen (see Table~\ref{tab:ARES_WPs}). Fig.~\ref{fig:LPSS_Sigma_ROI_Comp} shows the results. Starting from an overall linearity of the input LPS of $R=0.9568$, it can be seen that in all cases the use of the LPSS improved the linearity significantly. The smaller the region of interest (ROI) within the LPS, the better the results, reaching up to $R=0.999998$ in the case of $\tilde n = 1$. It is apparent that if the whole LPS is taken into account (i.e. a $6\,\sigma$ ROI), the results are noticeably worse than for a restricted ROI. This can be attributed to the fact that, due to the Gaussian time profile of the input LPS, the beam current in the head region of the bunch is low and hence the strength of the excited wakefield is weak. Thus, it is difficult for the optimizer to find configurations where this region is linearized sufficiently well, subsequently spoiling the overall linearity of the LPS. Excluding this head region of the LPS from the optimization, on the other hand, improves the performance significantly. In an experiment at ARES for example, the region outside of the ROI could be cut using the slit collimator implemented in the magnetic compressor (see below). 
In addition to the degree of linearity in the respective $\tilde n\sigma_z$ region, Fig.~\ref{fig:LPSS_Sigma_ROI_Comp} also shows that two important geometry parameters depend on the ROI as well. First, the overall structure length decreases with the ROI. This is of practical importance, not only in terms of beam transport through the structure, but also in terms of manufacturing. Second, the minimum aperture radius of the structure increases with a decreasing ROI, which is important from a beam transport point of view and in accordance with the time profile of the bunch and the dependence of the wake field strength per charge as shown in Fig.~\ref{fig:DLW_Modes_Geometry}. Taking these results into account, it is clearly worth considering trading - in case of a Gaussian time profile - less than \SI{5}{\percent} of the total bunch charge for the much better linearization performance of a $4\,\sigma$ ROI.

\begin{figure*}[htbp]
  \centering
  \includegraphics[width=\textwidth]{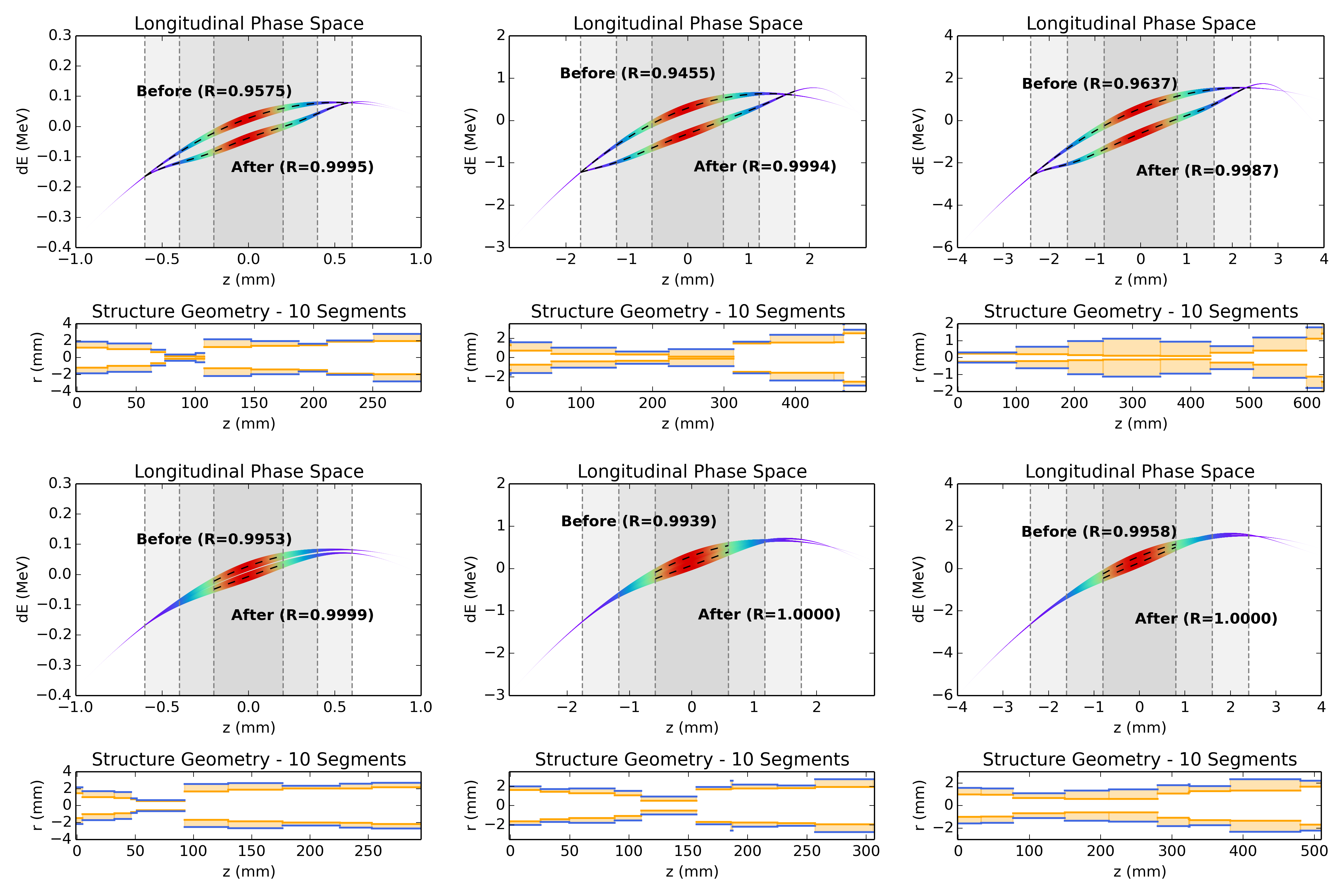}
  \caption{LPSS optimization results for input LPSs based on the ARES working points shown in Table~\ref{tab:ARES_WPs}. The optimization goal was to achieve $R=1$ across the full $6\,\sigma$ ROI (\textbf{top row}), as well as a centered $2\,\sigma$ ROI (\textbf{bottom row}). From left to right: WP1, WP2, WP3. Each plot shows the LPS before and after the LPSS interaction. The color and thickness visualize the current profile. The gray shaded areas correspond to the 2,4 and 6\,$\sigma$ regions respectively. The head of the bunch is on the left (negative $z$ values). Below the main plot, the geometry of the final segmented DLW is visualized, with the orange line corresponding to the inner radius and the blue line to the outer radius.}
  \label{fig:semiAna_Results_ARES_WPs}
\end{figure*}

Based on the results discussed above, optimization runs were performed for all of the three ARES working points, considering both a full $6\,\sigma$ and a smaller $2\,\sigma$ linearization ROI. Fig.~\ref{fig:semiAna_Results_ARES_WPs} shows the detailed results. It can be seen that in all cases a significant improvement of $R$ can be achieved within the ROI. Better results are obtained in the case of the limited ROI, as expected. Furthermore, the geometries of the resulting LPSS structures are shown. The shorter in time the input LPS, the shorter the resulting LPSS. This is partly due to the smaller required modulation depth, but also due to how the wakefield amplitude scales with the required inner radii of the segments ($E_z \propto 1/a^2$; see Fig.~\ref{fig:DLW_Modes_Geometry}). In order to accommodate a typical focused beam envelope, the individual segments of the LPSS structures are sorted such that the tightest segment is placed at the center of the structure, which then has increasing inner radii towards both entrance and exit. The results show that a similar degree of linearization can be achieved, regardless of the different bunch lengths across the different working points. The shape of the respective resulting structure does vary significantly however, due to the required modal content.
\subsection{Other Optimization Goals}
\begin{figure}[htbp]
	\centering
	\includegraphics[width=0.85\columnwidth]{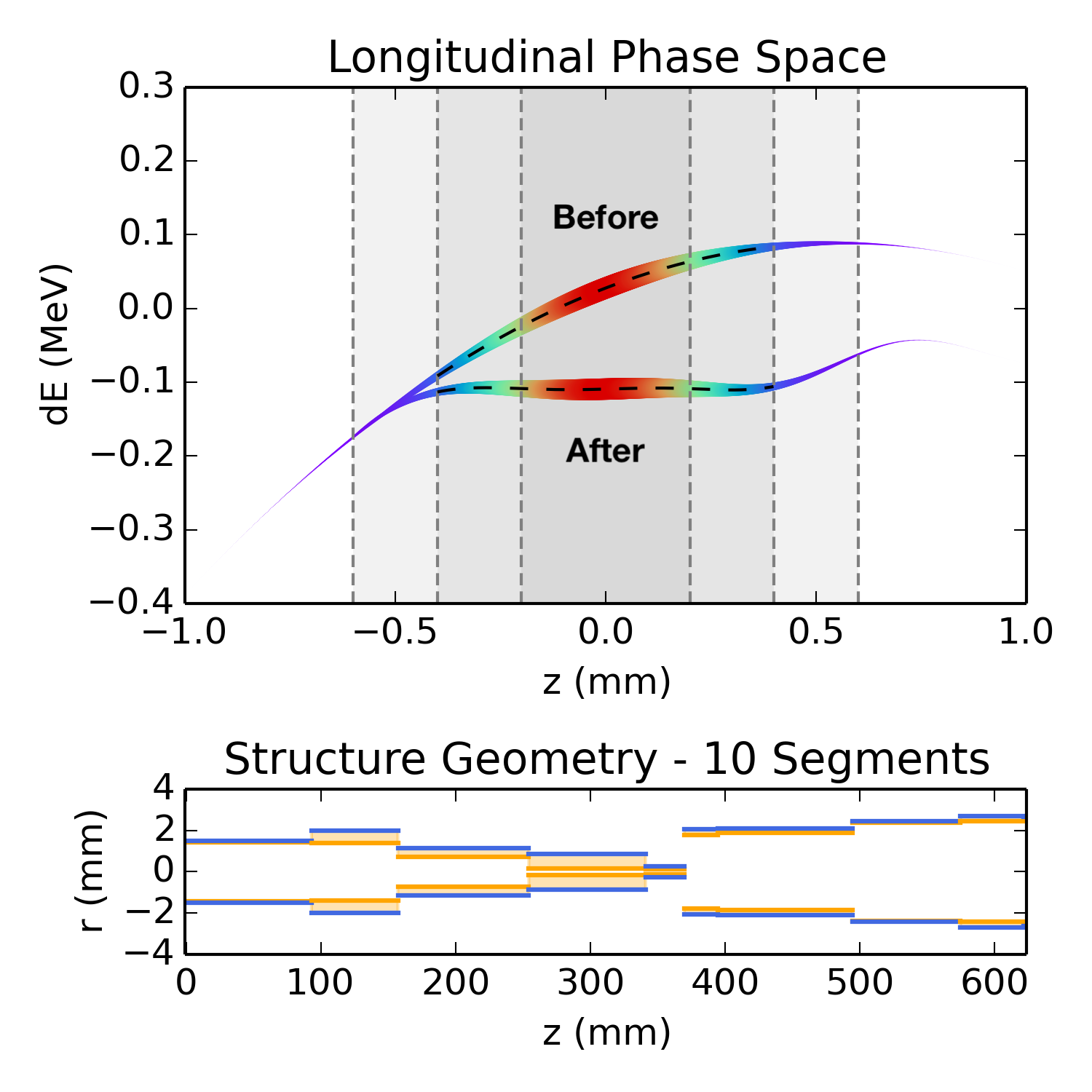}
	\caption{LPSS optimization results based on the ARES working point WP1 shown in Table~\ref{tab:ARES_WPs}. The optimization goal was to completely remove any correlated energy spread within a $4\,\sigma$ ROI. The color and thickness visualize the current profile. The gray shaded areas correspond to the 2,4 and 6\,$\sigma$ regions respectively. The head of the bunch is on the left (negative $z$ values). Below the main plot, the geometry of the final segmented DLW is visualized, with the orange line corresponding to the inner radius and the blue line to the outer radius.}
	\label{fig:semiAna_Results_ARES_WP1_Flat}
\end{figure}

As already discussed above, not only the linearization within a defined ROI can be set as an optimization goal. Another interesting case could be the removal of any correlated energy spread, aiming for a completely flat LPS. Fig.~\ref{fig:semiAna_Results_ARES_WP1_Flat} shows the result of such an optimization, based on the \SI{10}{\pico\coulomb} WP1 as shown in Table~\ref{tab:ARES_WPs}. It can be seen that the phase space is significantly flattened within the $4\,\sigma$ ROI. Note that this kind of structure could be used to prepare an LPS for further modulation as shown, for example, in section \ref{sec:Fourier}.
\subsection{Example Case: Bunch Compression}
\begin{figure}[htbp]
  \centering
  \includegraphics[width=0.86\columnwidth]{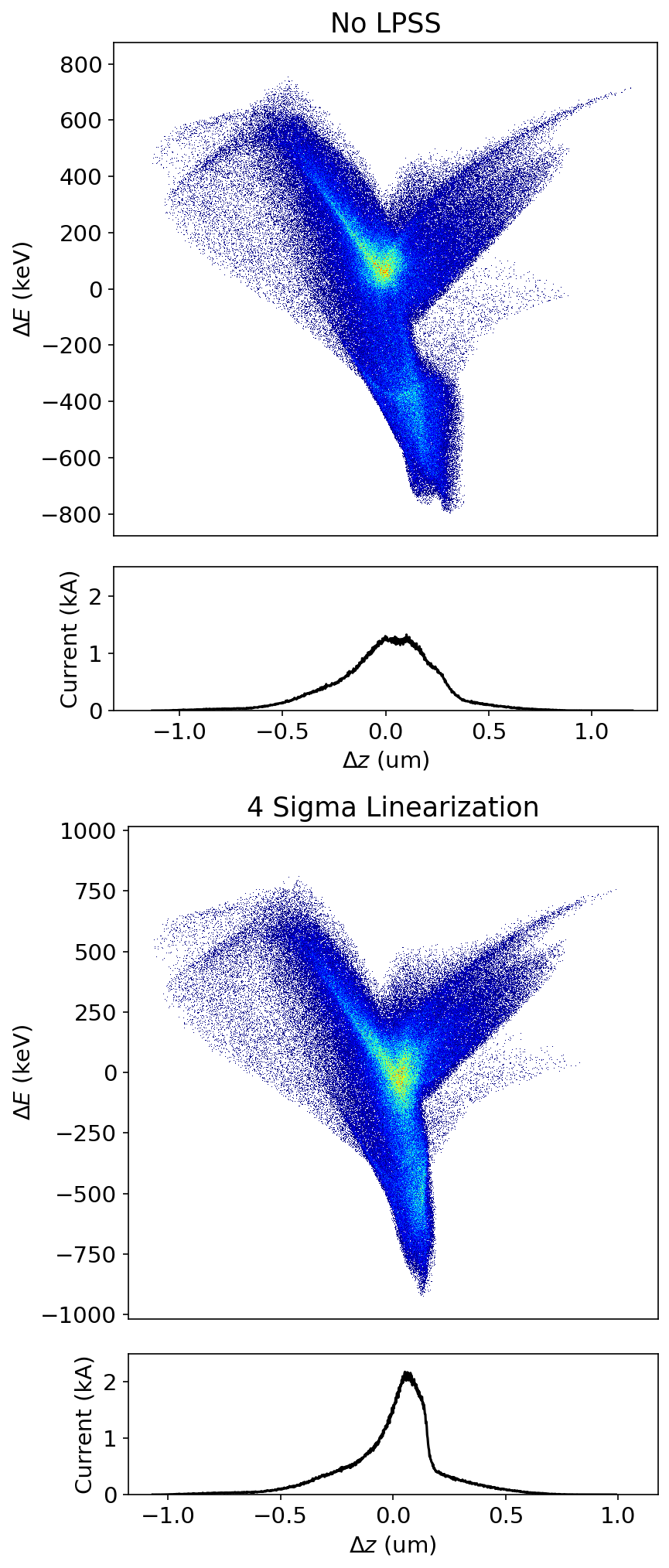}
  \caption{Numerical simulation of the longitudinal phase space and current profile of the ARES working point WP4 shown in Table~\ref{tab:ExampleWPs}, \SI{1.75}{\meter} downstream of the chicane exit ($z=\SI{30.5}{\meter}$). \textbf{Top:} Bunch compression without applying the LPSS optimization, i.e. no structure. \textbf{Bottom:} Bunch compression after applying a $4\,\sigma$ linearization with an optimized LPSS structure.}
  \label{fig:example_final_phase_spaces}
\end{figure}

It was shown in simulation that at ARES, based on magnetic compression and a slit-collimator, sub-fs bunch lengths can be achieved \cite{Zhu:2017phd, Zhu:2016ju}. Starting from an initial bunch charge of \SI{20}{\pico\coulomb} a final rms bunch length of \SI{0.51}{\femto\second} was achieved, \SI{1.75}{\meter} downstream of the chicane exit (cf. Fig.~\ref{fig:ARESLinac}). The remaining charge after the slit is \SI{0.79}{\pico\coulomb}, which corresponds to a $\sim \SI{4}{\percent}$ transmission. The full set of beam parameters is summarized in the first column of Table~\ref{tab:ExampleWPs}.
\begin{table}[!htbp]
   \centering
   \caption{Beam parameters of different ARES working points (WP) \SI{1.75}{\meter} downstream of the chicane exit ($z=\SI{30.5}{\meter}$). WP,Zhu taken from \cite{Zhu:2017phd}, WP4 obtained from ASTRA and IMPACT-T simulations, as well as the LPSS optimization routine. The $\tilde n\sigma$ subscript refers to the LPSS optimization ROI. TWS: Travelling Wave Structure.}
   \begin{ruledtabular}
   \begin{tabular}{lccc}
   \textbf{Parameter} & \textbf{WP,Zhu} & \textbf{WP4,$0\sigma$} & \textbf{WP4,$4\sigma$} \\
   \hline
   Initial charge & \SI{20}{\pico\coulomb} & \SI{10}{\pico\coulomb} & \SI{10}{\pico\coulomb}\\
   Final charge & \SI{0.79}{\pico\coulomb} & \SI{2.2}{\pico\coulomb} & \SI{2.18}{\pico\coulomb}\\
   TWS injection phase & \SI{-53}{\degree} & \SI{-38}{\degree} & \SI{-38}{\degree}\\
   Chicane $R_{56}$ & \SI{-12.4}{\milli\meter} &  \SI{-22.2}{\milli\meter} &  \SI{-22.2}{\milli\meter}\\
   Chicane slit width & \SI{0.4}{\milli\meter} &  \SI{0.6}{\milli\meter} &  \SI{0.6}{\milli\meter}\\
   $E_0$ & \SI{100.5}{\mega e\volt} &  \SI{126.0}{\mega e\volt} &  \SI{126.5}{\mega e\volt}\\
   $\sigma _E / E_0$ & $1.7 \cdot 10^{-3}$  &  $2.5 \cdot 10^{-3}$ &  $2.5 \cdot 10^{-3}$\\
   $\sigma _t$ & \SI{0.51}{\femto\second} & \SI{0.84}{\femto\second} & \SI{0.73}{\femto\second}\\
   $\varepsilon _{\text{n},x}$ & \SI{0.11}{\micro\meter} & \SI{0.35}{\micro\meter} & \SI{0.35}{\micro\meter}\\
   $\varepsilon _{\text{n},y}$ & \SI{0.1}{\micro\meter} & \SI{0.17}{\micro\meter} & \SI{0.13}{\micro\meter}\\
   $I _\text{p}$ & \SI{0.62}{\kilo\ampere} & \SI{1.32}{\kilo\ampere} & \SI{2.18}{\kilo\ampere}\\
   \end{tabular}
   \end{ruledtabular}
   \label{tab:ExampleWPs}
\end{table}
Here we aim to show that based on using an LPSS before magnetic bunch compression, we can achieve similar beam parameters, but at higher mean energy and higher final peak current. To this end WP4, which is a modified version of WP1 (cf. Table~\ref{tab:ARES_WPs}), where the TWS structures are driven at \SI{-38}{\degree} is used in a start-to-end simulation using ASTRA, the LPSS optimization routine and IMPACT-T \cite{IMPACT-T:2006prab}. Up to the LPSS structure the simulation includes space charge forces via ASTRA and after that both space charge and CSR via IMPACT-T. Full linearization in a $4\,\sigma$ optimization ROI was considered as the LPSS optimization goal. The resulting beam parameters \SI{1.75}{\meter} downstream of the chicane exit are summarized in Table~\ref{tab:ExampleWPs}, where WP4,$0\sigma$ refers to our working point without LPSS linearization and WP4,$4\sigma$ to the case employing the optimized LPSS structure. The final longitudinal phase spaces are shown in Fig.~\ref{fig:example_final_phase_spaces}. It can be seen that using a passive LPSS structure upstream of the magnetic bunch compressor in $\tilde n\sigma$ linearization mode yields bunches with similar beam quality, but at \SI{26}{\percent} higher mean energy. At the same time, even though the initial charge is \SI{50}{\percent} less, the final charge is higher, due to the larger slit width. This is possible due to the high degree of linearization in the LPSS ROI. The peak current is noticeably higher in both WP4 cases ($\sim 2\times$ w.o. the LPSS and $\sim 3.5\times$ using the LPSS). 

We note that the transverse phase space of WP4 was not fully optimized as part of this study, which means that the transverse properties of the beam could be improved in future iterations of this particular working point.

Finally it should be noted, that at higher overall charges significant energy modulation due to CSR can spoil the linearity of the LPS during bunch compression. This, however, could be included into future versions of the LPSS optimization routine as the virtual last element of the LPSS structure.
\section{Realistic Structures} \label{sec:RealisticStructures}
\subsection{Segment Transitions}
\begin{figure}[htbp]
  \centering
  \includegraphics[width=\columnwidth]{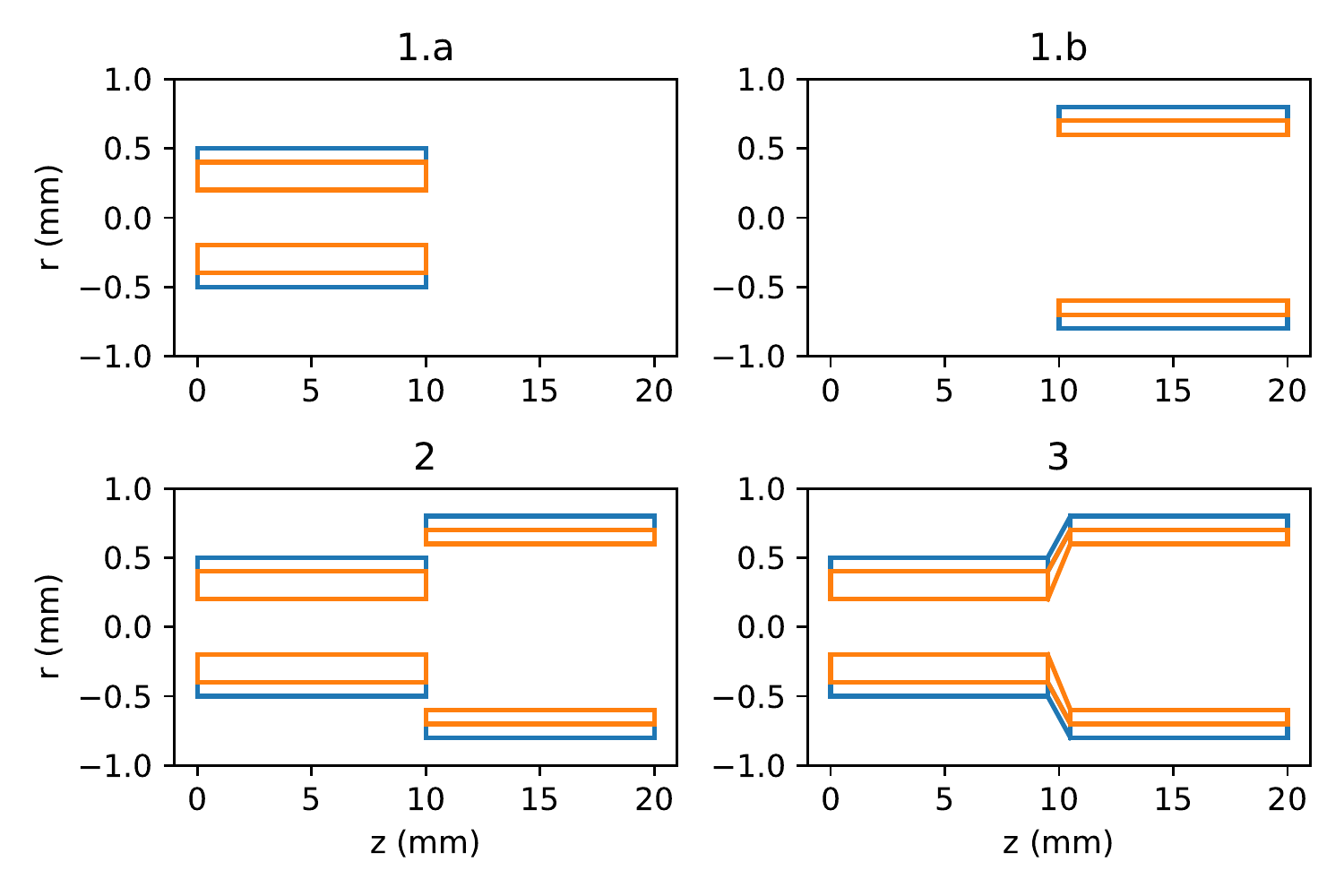}
  \caption{Illustration of the DLW geometry used in the ECHO2D simulations. All cases include a (lossless) metal coating of \SI{100}{\micro\meter} thickness. The blue lines correspond to the outline of the metal coating and the orange lines to the outline of the dielectric. \textbf{1.a}: Single segment of length $l_1 = \SI{10}{\milli\meter}$. \textbf{1.b}: Single segment of length $l_2 = \SI{10}{\milli\meter}$. \textbf{2}: Segments right next to each other (sharp, unrealistic transition). \textbf{3}: Two segments connected with a linearly tapered transition region of length $l_t = \SI{1}{\milli\meter}$.}
  \label{fig:ECHO_SIM_1}
\end{figure}

Our previous discussion has treated the LPSS as a series of individual successive DLW segments. In order to calculate the resulting energy modulation, the individual wakefields of the segments were summed up and applied to the input LPS. Although this is a good first approximation, in reality there are two issues with this approach. First, the sharp transitions between the segments will disturb the wakefield slightly. Second and most importantly, this kind of segmented structure cannot be produced, because in some cases it turns out that $a_{i+1} > b_i$, which would mean that the ($i+1$)th segment could not actually be attached to the $i$th segment. It is hence necessary to include transition elements between the individual segments. These elements could for example be short linearly tapered sections. Although adding such a transition would enable production of the structure, it also alters the resulting wakefield. In order to investigate this effect, ECHO2D \cite{ECHO2D:33} simulations were performed. The longitudinal monopole wakefield, excited by a Gaussian current with an arbitrarily chosen $\sigma _t = \SI{500}{\femto\second}$, was compared for three different cases:
\begin{enumerate}
    \item The sum of the resulting wakefield of two individually simulated DLW segments of length $l_1$ and $l_2$,
    \item The two segments directly behind one another. (sharp, unrealistic transition),
    \item The two segments connected with a linearly tapered transition region of length $l_t$.
\end{enumerate}
Note that the overall length $L$ of the structure is the same for both case 2 and 3. This means that for case 3 the individual segments are shortened by $0.5\cdot l_t$ each. Hence, case 2 is essentially case 3 with $l_t = 0$. See Fig.~\ref{fig:ECHO_SIM_1} for an illustration of the three different cases.

\begin{figure}[hbp]
  \centering
  \includegraphics[width=\columnwidth]{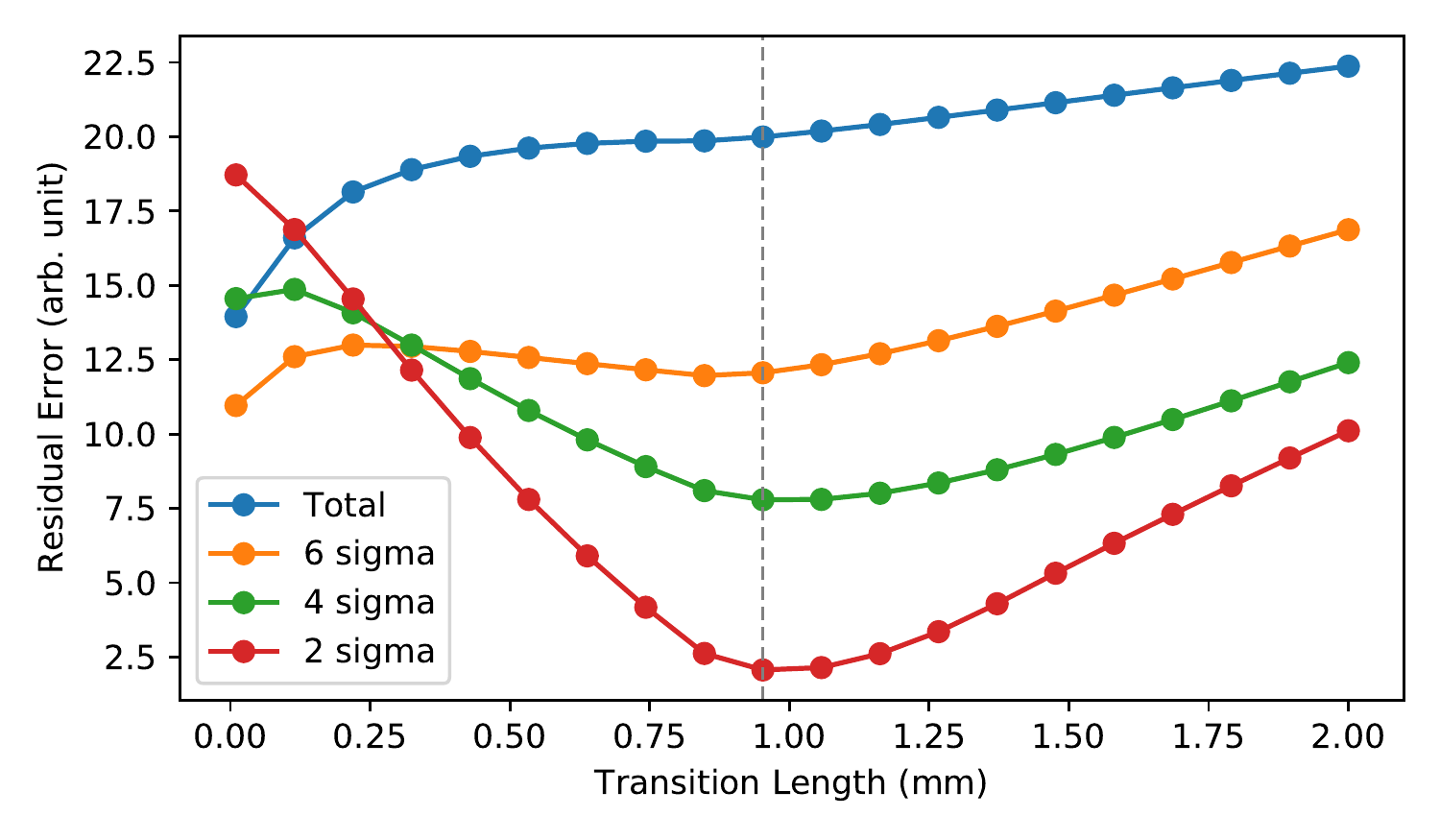}
  \caption{Normalized integrated residual difference between the wakefield obtained from the sum of two singular DLW segments and a combined device with a linearly tapered transition region of length $l_t$, as show in Fig.~\ref{fig:ECHO_SIM_1}. The different curves correspond to the $6\sigma, 4 \sigma$ and $2 \sigma$ parts of the drive bunch, as well as the complete simulation box (total).}
  \label{fig:ECHO_SIM_2}
\end{figure}

Fig.~\ref{fig:ECHO_SIM_2} shows the integrated residual difference between the wakefield obtained from the case 1 and case 3 geometries using a drive bunch with $\sigma _t = \SI{500}{\femto\second}$ vs. different values of $l_t$. The exemplary dimensions of the DLW segments are $a_1 = \SI{0.2}{\milli\meter}$, $b_1 = \SI{0.4}{\milli\meter}$, $l_1 = \SI{10}{\milli\meter}$, $a_2 = \SI{0.6}{\milli\meter}$, $b_2 = \SI{0.7}{\milli\meter}$, $l_2 = \SI{10}{\milli\meter}$. The dielectric is defined by $\epsilon _r = 4.41, \mu _r = 1$ and the metal coating, which is assumed to be a perfect conductor, has a thickness of \SI{0.1}{\milli\meter}. The simulation results show that an optimal $l_t$ can be found depending on the area of interest around the peak of the drive current. It has to be noted that although this minimum does not depend strongly on the longitudinal dimensions of the segments, it does depend on the transverse dimensions $a_i$ and $b_i$ (and on $\sigma _t$, as the whole composition of the structure depends on it). It is hence implied that each transition has to be uniquely optimized. This, however, can be directly factored into the optimization routine discussed above (extending the number of degrees of freedom from $3N$ to $4N-1$).

\begin{figure}[htbp]
  \centering
  \includegraphics[width=\columnwidth]{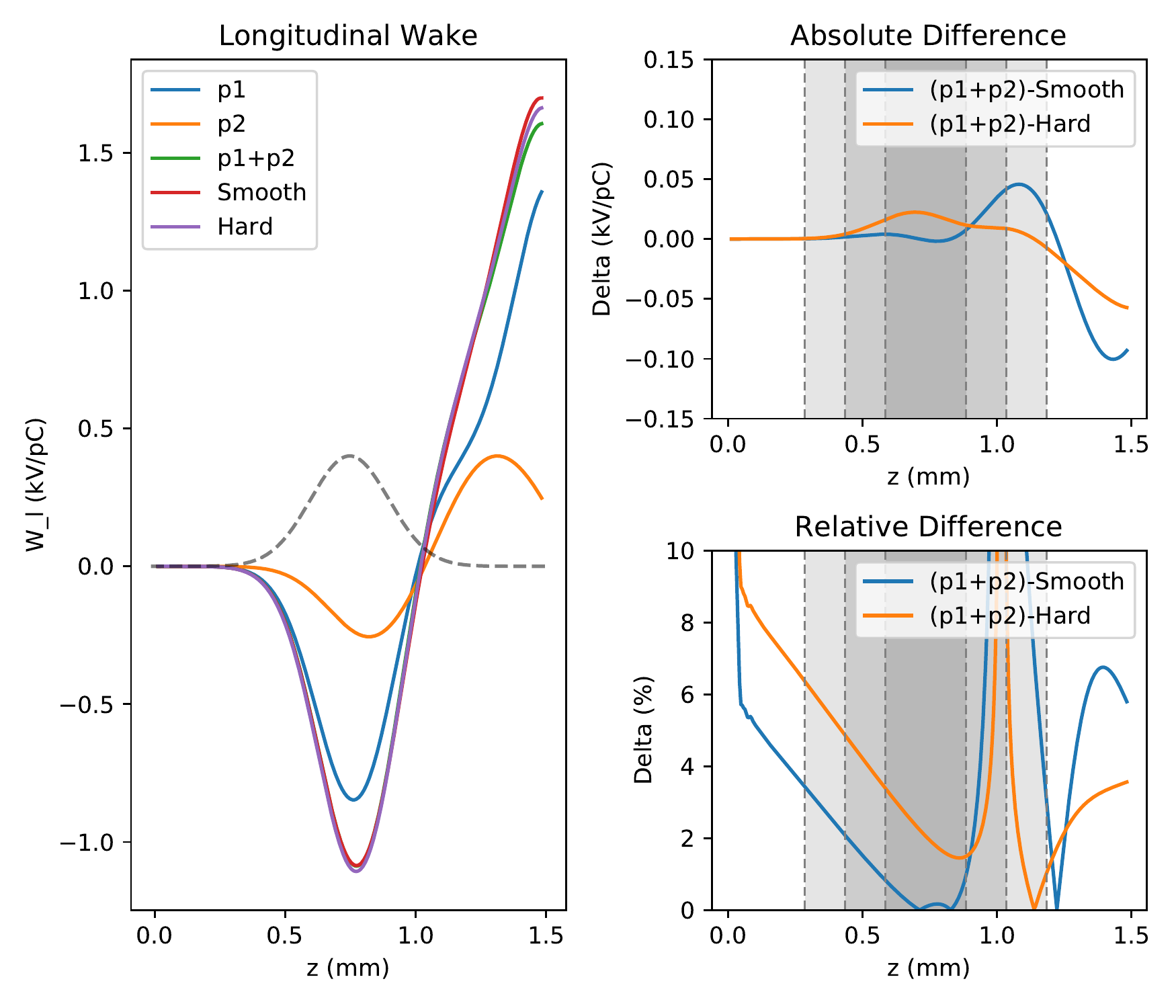}
  \caption{Comparison of the wakefield obtained using the geometries illustrated in Fig.~\ref{fig:ECHO_SIM_1}. $l_t = \SI{953}{\micro\meter}$, which is the value determined by the optimization scan shown in Fig.~\ref{fig:ECHO_SIM_2}. The shaded areas correspond to the $6\sigma, 4 \sigma$ and $2 \sigma$ parts of the drive bunch.}
  \label{fig:ECHO_SIM_3}
\end{figure}

It was shown that the integrated difference between a case 1 and 3 geometry can be minimized by adjusting $l_t$. Fig.~\ref{fig:ECHO_SIM_3} shows the longitudinal wake for all three geometry cases based on a simulation using the exemplary parameters from above and an optimized $l_t$ of \SI{953}{\micro\meter}. In addition to the wakefields, the absolute and relative difference compared to case 1 is plotted for both the case 2 and 3 geometry respectively. It can be seen that, depending on the area of interest along the drive bunch, the error can be very small and is generally smaller than \SI{10}{\percent}. The error can be large, however, towards the tail of the bunch. The significance of this effect depends a lot on the specific input electron distribution and the particular use case. Assuming a Gaussian longitudinal current profile, $< \SI{16}{\percent}$ of the charge is affected. Recalling Fig.~\ref{fig:ECHO_SIM_2}, the goal should in general be to minimize the effect of the transition in the region of highest charge density. In summary, it can be concluded that it is possible to find transition regions, which minimize the difference of the produced wakefield compared to the summed up wakefield of individual segments, as used in the optimization routine discussed above.
\subsection{Manufacturing}
\begin{figure}[htbp]
  \centering
  \includegraphics[width=\columnwidth]{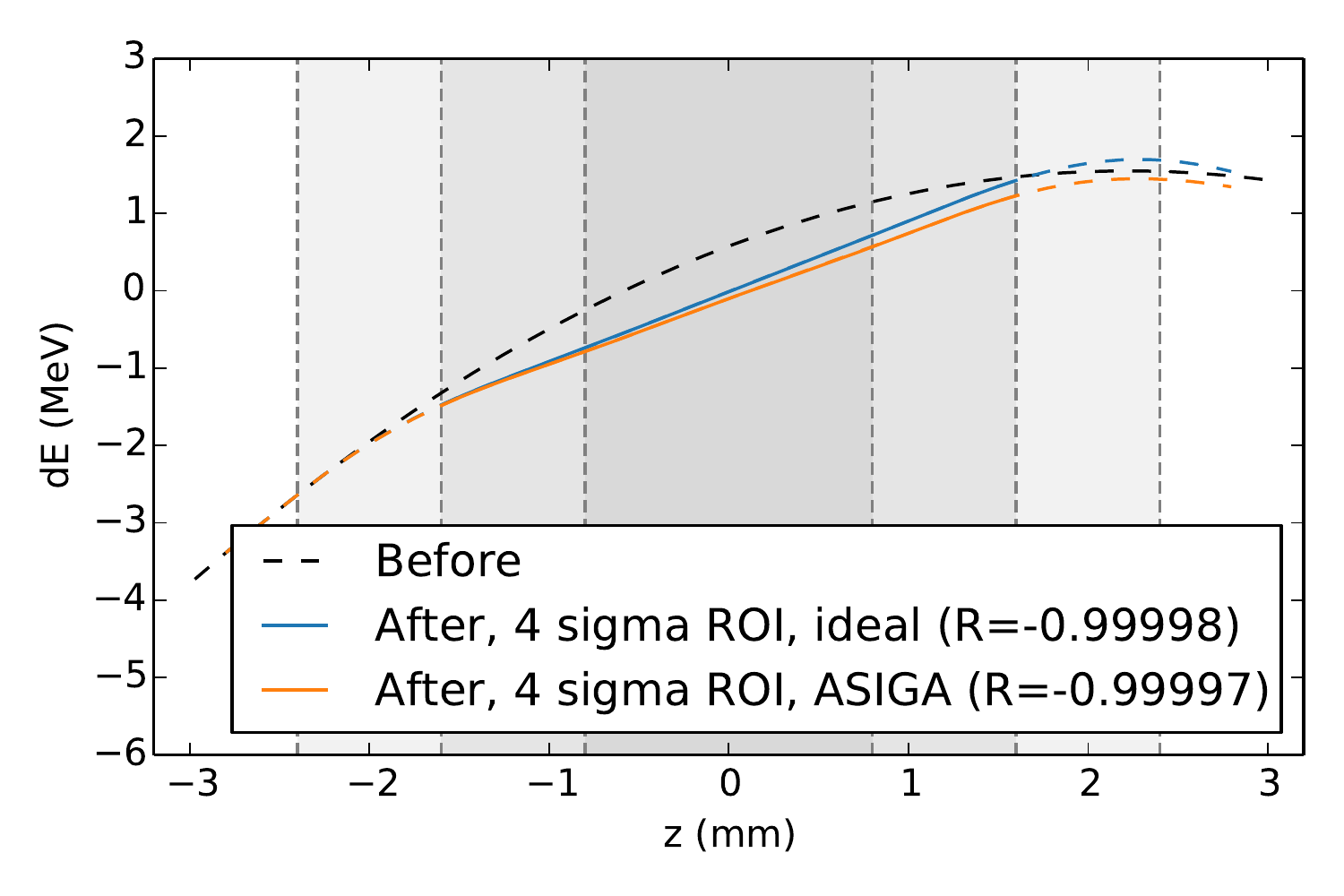}
  \caption{Comparison of LPSS optimization results for a Gaussian input current profile. The optimization goal was to achieve $R=1$ in a 4\,$\sigma$ region of interest. The input beam parameters correspond to WP3 (see Table~\ref{tab:ARES_WPs}). \textbf{Blue:} Ideal LPSS, \textbf{orange:} LPSS taking a lateral printing resolution of \SI{27}{\micro\meter} and a longitudinal printing resolution of \SI{10}{\micro\meter} into account (as can, for example, be achieved with an ASIGA MAX X27 3D printer).}
  \label{fig:semiAna_Result_Gauss_ASIGA}
\end{figure}

The optimization shown above does not include any assumption about possible inaccuracies due to the manufacturing process. In reality, the exact shape of the individual segments is determined by the tolerances during production. Assuming a 3D-printed structure, the parameters $a_i$, $b_i$ and $l_i$ are determined by the transverse and longitudinal printing resolution and on how the structure is printed (flat or standing). We consider the ASIGA MAX X27 3D printer \cite{ASIGA:MAXX27} and its printing resolution as an example. This particular printer has a longitudinal resolution $\rho_{z}$ of \SI{10}{\micro\meter} (minimum layer thickness) and a lateral resolution $\rho_{xy}$ of \SI{27}{\micro\meter} (DLP pixel size). Fig.~\ref{fig:semiAna_Result_Gauss_ASIGA} shows the comparison between the linearization using an ideal LPSS and an LPSS, which was optimized taking the aforementioned printing resolution into account. Here we model the effect such that $\tilde a_i = \lfloor 2a_i/\rho_{xy} \rfloor \cdot \rho_{xy}/2$, $\tilde b_i = \lceil 2b_i/\rho_{xy} \rceil \cdot \rho_{xy}/2$ and $\tilde l_i = \lceil l_i/\rho_{z} \rceil \cdot \rho_{z}$,
where the tilde denotes the radii and length of the segments after applying the printer resolution. The results show that the limited printing resolution only has a small impact on the final linearization. It has to be noted, that the chirp across the ROI is different, but only because it was not part of the particular optimization goal.
\section{Robustness of the Scheme} \label{sec:robustness}
\subsection{Input LPS}
\begin{figure}[htbp]
  \centering
  \includegraphics[width=\columnwidth]{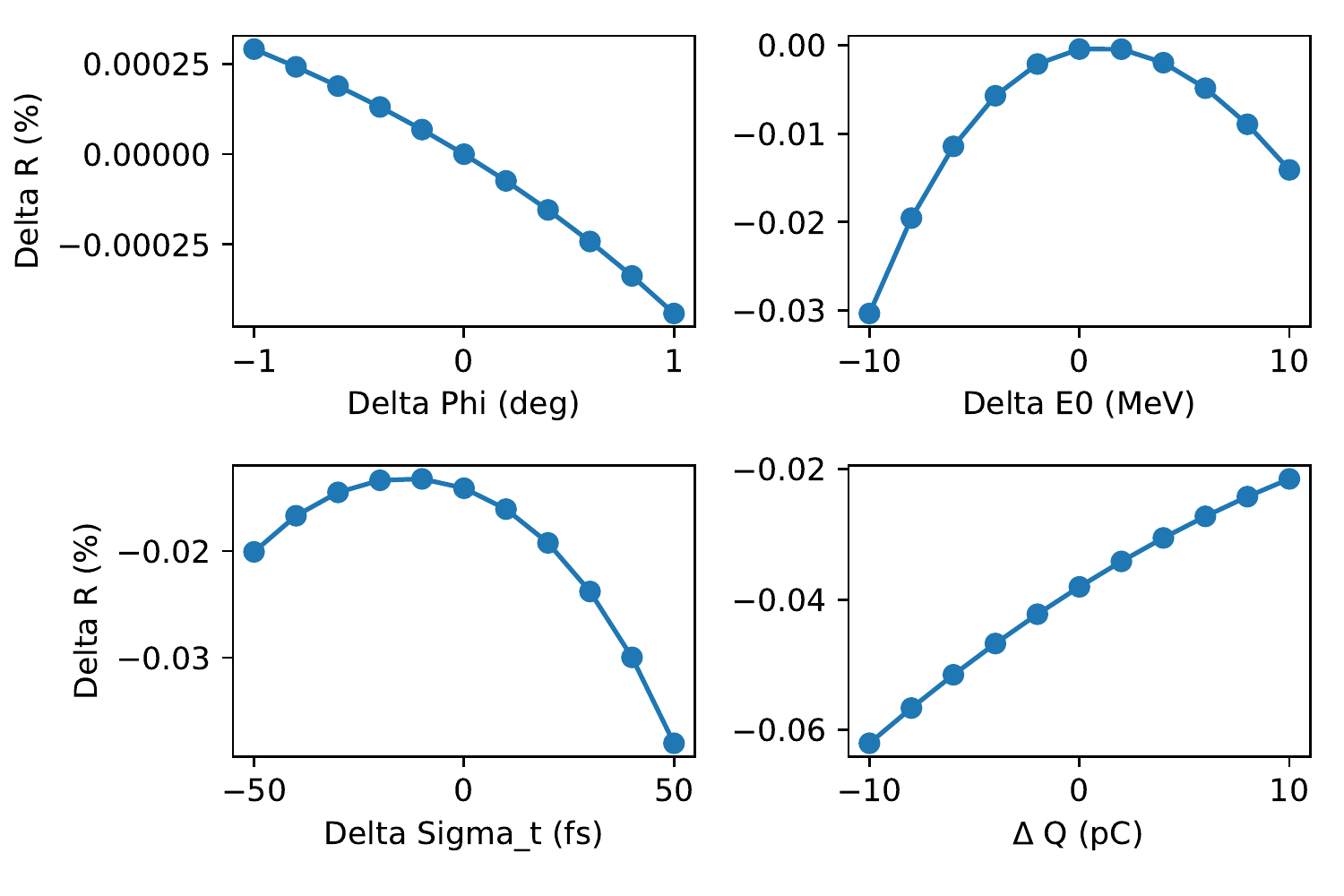}
  \caption{Relative change of the linearity factor $R$ vs. four different parameters, which influence the input LPS.}
  \label{fig:Robustness_Results}
\end{figure}

As discussed above, an LPSS must be specifically tailored to the incoming LPS. In reality the actual shape of the input LPS varies according to the stability of certain accelerator machine parameters. The LPS in particular is influenced by the stability of both amplitude and phase of the accelerating fields, but also by dispersive sections and collective effects, such as coherent synchrotron radiation (CSR). It is hence interesting to investigate the effect of the actual shape of the input LPS on the output LPS. To this end, the third ARES working point (WP3) with \SI{200}{\pico\coulomb} of total charge and a Gaussian time profile with $\sigma_t = \SI{2.65}{\pico\second}$ (see Section~\ref{sec:Multimode_Example}) is used. The sensitivity of the linearity parameter $R$ within a 4\,$\sigma$ ROI is determined for four different parameters, with the first two parameters being the amplitude and phase of the accelerating field, which define the curvature of the incoming LPS. The third parameter is $\sigma _t$, which in reality, of course, non-trivially depends on multiple factors, but is here varied independently, while keeping the total bunch charge constant. The fourth parameter is the bunch charge $Q$, keeping $\sigma _t$ constant. Fig.~\ref{fig:Robustness_Results} summarizes the results of the four scans. The results show that the relative change in $R$ is very little ($\ll \SI{0.1}{\percent}$), leading to the conclusion that, in the specific case of the example of LPS linearization, the LPSS scheme is robust within the limits of typical accelerator machine stability.
\subsection{Systematic Manufacturing Errors}
\begin{figure}[htbp]
  \centering
  \includegraphics[width=\columnwidth]{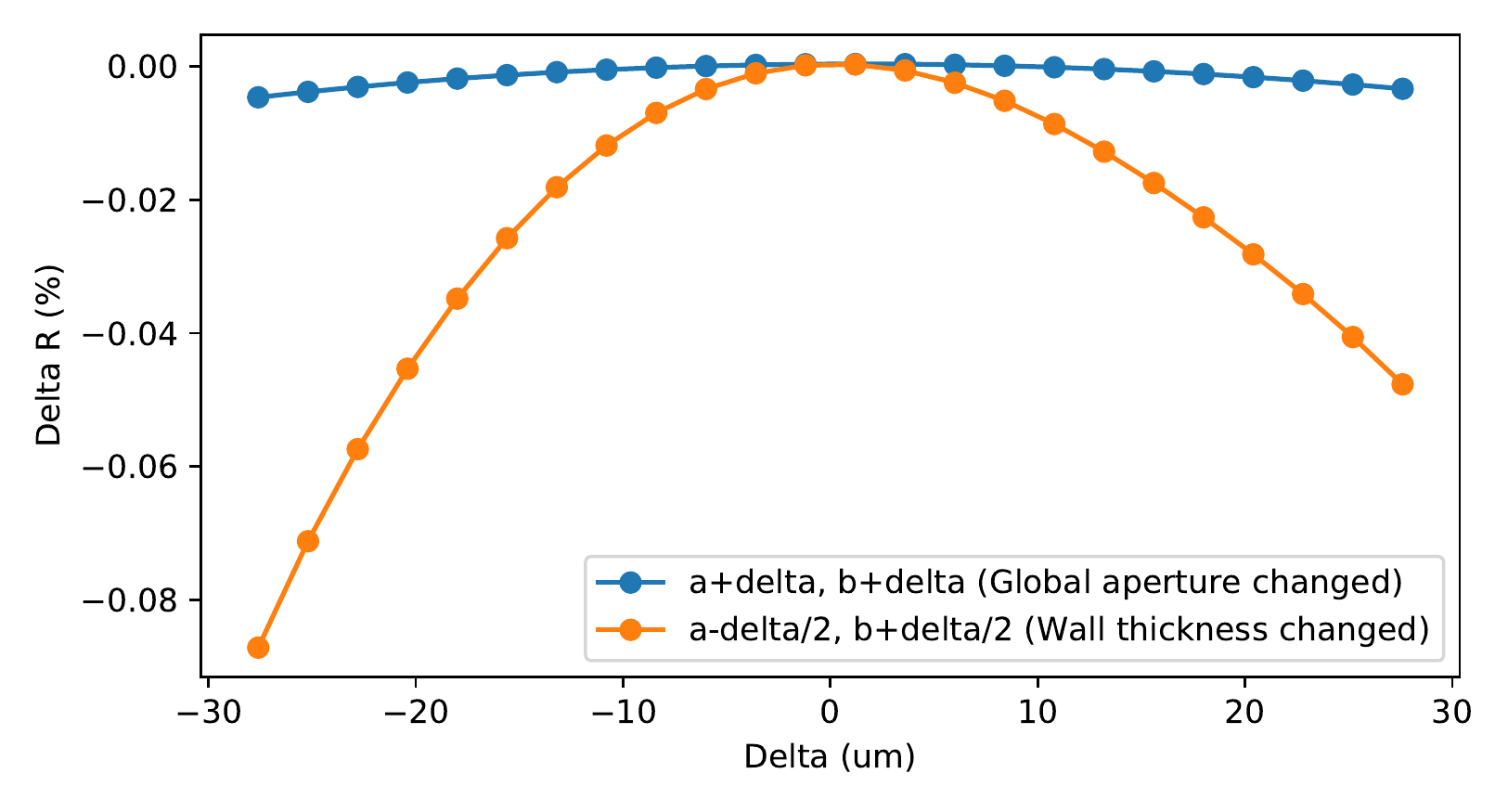}
  \caption{Relative change of the linearity factor $R$ vs. a constant error $\Delta r$ for two different systematic error scenarios. The same $\Delta r$ is applied to all segments.}
  \label{fig:Robustness_Results_ASIGA}
\end{figure}

In addition to the uncertainty in the shape of the input LPS, there can also be systematic errors in the geometry of the LPSS itself. In order to investigate this, two scenarios were studied. The first one is a constant error $\Delta r$ of both the inner and outer radii, i.e. $\tilde a_i = a_i+\Delta r$ and $\tilde b_i = b_i+\Delta r$. The second scenario is a constant difference in wall thickness, meaning $\tilde a_i = a_i-\Delta r/2$ and $\tilde b_i = b_i+\Delta r/2$. The range is chosen to be according to the lateral resolution of the ASIGA printer discussed above. Hence $\Delta r \in [-30,30]\,\SI{}{\micro\meter}$. Fig.~\ref{fig:Robustness_Results_ASIGA} summarizes the results of the scan. The LPSS optimization scenario is the same as before. It can be seen that the change in global aperture has a very small effect on $R$ ($< \SI{0.01}{\percent}$). The wall thickness, on the other hand, has a $\sim 10\times$ stronger effect, with a slight asymmetry. It is still a small effect with $|\Delta R| < \SI{0.1}{\percent}$ within the given range of $\Delta r$. The slightly asymmetric behaviour might be explained by the non-linear dependence of the amplitude and frequency of the wake towards smaller inner radii (cf. Fig.~\ref{fig:DLW_Modes_Geometry}) in conjunction with an increase in the modal content as the thickness of the dielectric lining increases. A more thorough study of this behaviour would be interesting, but exceeds the scope of this work. 
\section{Conclusion and Outlook}
A completely passive LPSS solution, based on segmented DLWs was presented and studied both analytically and numerically. The results based on the idealized single-mode Fourier synthesis, coupled with a longitudinally dispersive section, reveal phase space configurations, which could be interesting for applications, especially in the context of radiation generation (multi-color microbunch trains, sub-microbunches with tunable relative spacing, etc.).

Arbitrary multimode optimization was investigated, which enables application of the method to arbitrary input phase spaces. The results shown here are promising, as the exemplary goal of full linearization of the input LPS within a given $\tilde n\sigma_z$ ROI of an input LPS was achieved in a semi-analytical simulation to a very high degree. The input LPS used for the study were chosen to be realistic and are based on numerical simulation of an existing accelerator, the ARES linac at DESY. Motivated by these results, a start-to-end simulation of a possible experiment at the ARES linac was performed yielding sub-fs bunches comparable to reference working points, but at $\sim \SI{26}{\percent}$ higher mean energy and $\sim 3.5\times$ larger peak current, starting from \SI{50}{\percent} less initial charge. 

\begin{figure}[htbp]
	\centering
	\includegraphics[width=0.4\columnwidth]{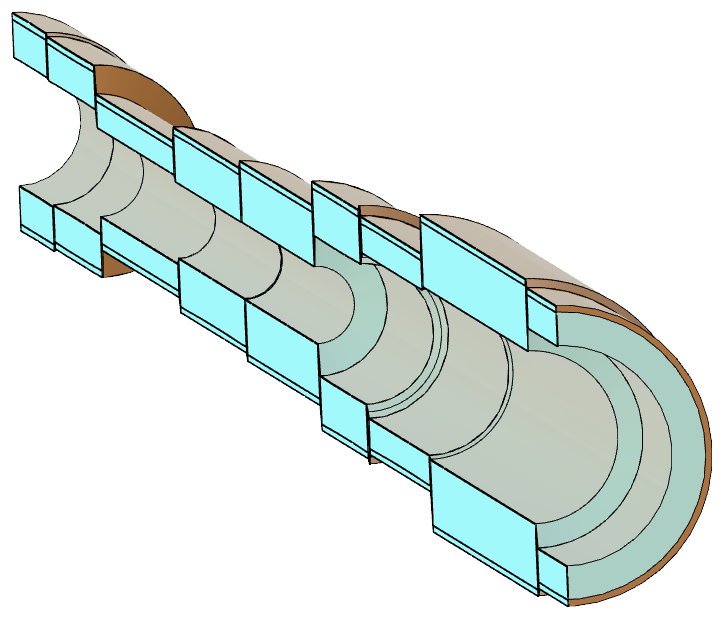}
	\caption{Section view of a 3D rendering of a potential printed and metallized LPSS structure. The 3D model was obtained directly from the optimization routine.}
	\label{fig:3DRendering}
\end{figure}

It was furthermore shown, based on ECHO2D simulations, that it is possible to integrate short transition regions between the segments, which enables realistic structure shapes that can be produced with a 3D-printer. The optimization routine used in this work can export its result as 3D models suitable for direct import into a 3D printing software. Fig.~\ref{fig:3DRendering} shows a rendering of such a file. The structures can be made from metallized 3D-printed plastic, or even 3D-printed quartz \cite{Kotz:2017ez}. Depending on the specific printing process, longer structures might be constructed of two or more cascaded macro segments. 

The robustness of the scheme was investigated for the LPS linearization example and found to be satisfactory based on accelerator stability, as well as manufacturing tolerance considerations. This together with the low cost of the devices alleviates the fact that each LPSS device is specific to a given accelerator working point; multiple structures could be installed and swapped in as needed. 

Further studies could focus on transverse effects in LPSS structures, as potentially triggered dipole modes might lead to deflection. Also material-dependent charging of the dielectric could be studied. Finally, the LPSS optimization routine could be updated to take expected downstream LPS modulation, due to e.g. collective effects, into account.
\begin{acknowledgments}
The authors would like to thank I.~Zagorodnov for support in the use of ECHO2D.
\end{acknowledgments}
\bibliography{LPSS}
\end{document}